\documentclass[twocolumn,showpacs,amsmath,amssymb,prl,superscriptaddress]{revtex4-1}
\usepackage{graphicx}
\usepackage{dcolumn}
\usepackage{bm}
\usepackage{color}%
\usepackage{comment}
\begin{document}

\title{High temperature antiferromagnetism in Yb based heavy fermion systems
proximate to a Kondo insulator}

\author{Shintaro Suzuki}
\thanks{These authors contributed equally to this work.}
\affiliation{Institute for Solid State Physics (ISSP), University of Tokyo, Kashiwa, Chiba 277-8581, Japan}
\author{Kou Takubo}
\thanks{These authors contributed equally to this work.}
\affiliation{Institute for Solid State Physics (ISSP), University of Tokyo, Kashiwa, Chiba 277-8581, Japan}
\author{Kentaro Kuga}
\affiliation{RIKEN SPring-8 Center, Sayo, Hyogo 679-5148, Japan}
\author{Wataru Higemoto}
\affiliation{Advanced Science Research Center, Japan Atomic Energy Agency, Tokai, Ibaraki 319-1195, Japan}
\affiliation{Department of Physics, Tokyo Institute of Technology, Meguro, Tokyo 152-8550, Japan}
\author{Takashi U. Ito}
\affiliation{Advanced Science Research Center, Japan Atomic Energy Agency, Tokai, Ibaraki 319-1195, Japan}
\author{Takahiro Tomita}
\affiliation{Institute for Solid State Physics (ISSP), University of Tokyo, Kashiwa, Chiba 277-8581, Japan}
\author{Yasuyuki Shimura}
\affiliation{Graduate School of Advanced Science and Engineering, Hiroshima University, Higashi Hiroshima, 739-8530, Japan}
\author{Yosuke Matsumoto}
\affiliation{Max Planck Institute for Solid State Research, Heisenbergstrasse 1, Stuttgart 70569, Germany}
\author{C\'{e}dric Bareille}
\affiliation{Institute for Solid State Physics (ISSP), University of Tokyo, Kashiwa, Chiba 277-8581, Japan}
\author{Hiroki Wadati}
\affiliation{Institute for Solid State Physics (ISSP), University of Tokyo, Kashiwa, Chiba 277-8581, Japan}
\author{Shik Shin}
\affiliation{Institute for Solid State Physics (ISSP), University of Tokyo, Kashiwa, Chiba 277-8581, Japan}
\author{Satoru Nakatsuji}
\thanks{To whom correspondence should be addressed.}
\email{satoru@phys.s.u-tokyo.ac.jp}
\affiliation{Institute for Solid State Physics (ISSP), University of Tokyo, Kashiwa, Chiba 277-8581, Japan}
\affiliation{Department of Physics, Graduate School of Science, University of Tokyo, Tokyo 113-0033, Japan}
\affiliation{CREST, Japan Science and Technology Agency (JST), 4-1-8 Honcho Kawaguchi, Saitama 332-0012, Japan}
\affiliation{Institute for Quantum Matter and Department of Physics and Astronomy, Johns Hopkins University, Baltimore, Maryland 21218, USA}
\affiliation{Trans-scale Quantum Science Institute, University of Tokyo 113-0033, Tokyo, Japan}

\date{\today}

\begin{abstract}
Given the parallelism between the physical properties of Ce and Yb based
magnets and heavy fermions due to the electron-hole symmetry, it has been
rather odd that the transition temperature of the Yb based compounds is
normally very small, as low as $\sim$ 1 K or even lower, whereas Ce
counterparts may often have the transition temperature well exceeding 10 K.
Here, we report our experimental discovery of the transition temperature
reaching 20 K for the first time in a Yb based compound at ambient pressure.
The Mn substitution at the Al site in an intermediate valence state
of $\alpha$-YbAlB$_{4}$ not only induces antiferromagnetic transition at a
record high temperature of 20 K but also transforms the heavy fermion liquid
state in $\alpha$-YbAlB$_{4}$ into a highly resistive metallic state proximate
to a Kondo insulator.

\end{abstract}

\maketitle

Correlated electron systems have provided a number of non-trivial phenomena
including quantum criticality, unconventional superconductivity
and exotic magnetic/electronic orders.
Specifically, $4f$ electron systems have provided prototypical materials
to study quantum criticality that appears as a result of the competition
between the RKKY interaction and the Kondo effect \cite{RevModPhys.79.1015,
gegenwart2008quantum,PhysRevLett.67.3310,PhysRevLett.84.4986,
gegenwart2002magnetic,yuan2003observation,Bianca2003possible,
nakatsuji2008superconductivity,sakai2011kondo,matsubayashi2012pressure,
tsujimoto2014heavy}.
$4f$ electron systems also provide ideal platforms to investigate the effects
of correlation in the presence of the strong spin-orbit coupling.
The discovery of novel topological phases in 
correlated electron systems such as topological insulators and Weyl semimetals
progressively attracted attention to even more strongly correlated
system~\cite{hasan2010colloquium,huang2015weyl,Xu613,PhysRevX.5.031013,
nakatsuji2015large,kuroda2017evidence},
including 4$f$ electron systems with, for example, the so-called
topological Kondo insulators~\cite{dzero2010topological,wolgast2013low,
li2014two,neupane2013surface}.

Among all lanthanide ions, most studies have focused on Ce and Yb based
compounds as they are at the one electron and the one hole limits of
the 4$f$ shell respectively.
This electron-hole symmetry implies a parallelism between the physical
properties of Ce and Yb based magnets and heavy fermions.
However, the transition temperature of the Yb based compounds is normally very
small,
as low as $\sim$ 1 K or even lower \cite{gegenwart2002magnetic,
bauer1997magnetic,AVILA200356,1742-6596-273-1-012048,PhysRevB.35.1914,
PhysRevLett.67.3310,DHAR1999150,PhysRevB.60.1136,PhysRevB.69.014415,
BENNETT20092021,POLLIT1985583,SCHANK19951237},
whereas Ce counterparts may have the transition temperature often well
exceeding 10 K \cite{BENOIT1980293,PhysRevB.16.440,knebel2009high,
nishioka2009novel,kawabata2014suppression}.

Among a number of Yb-based heavy fermion systems, $\beta$-YbAlB$_{4}$ is
particularly interesting as it is the first example of a heavy fermion
superconductor and unconventional quantum criticality without tuning, namely,
at ambient pressure and at zero field \cite{nakatsuji2008superconductivity,
kuga2008superconducting,matsumoto2011quantum,tomita2015strange}.
In contrast, its isomorphic compound $\alpha$-YbAlB$_{4}$ has a Fermi liquid
ground state while it exhibits almost the same Kondo lattice like behavior
at higher temperatures \cite{matsumoto2011anisotropic}.
Recent study revealed that a small amount of Fe doping of 1.4\% induces nearly
the same type of quantum criticality as found in $\beta$-YbAlB$_{4}$
at ambient pressure \cite{Kuga2017}.
The clear anomaly in the low temperature valence indicates quantum valence
transition as the most likely
origin \cite{watanabe2010quantum,Kuga2017,PhysRevLett.104.247201}.

Another particularly striking feature found in these compounds is the unusually
high transition temperature of their magnetism.
For example, the transition temperature exceeding 30 K was observed for
$\beta$-YbAlB$_{4}$ under high pressure, which is a record high transition
temperature for Yb based compounds to date \cite{tomita2015high,
tomita2016pressure}.
In addition, for both $\alpha$ and $\beta$ phases of YbAlB$_{4}$,
a sufficient Fe doping stabilizes an antiferromagnetic ordering whose
transition temperature becomes as high as 10 K at ambient
pressure~\cite{kuga2012magnetic,kuga2014two,Kuga2017}.
These fascinating observations of the unusually high transition temperature
as well as the unconventional quantum criticality in these systems
indicate a novel mechanism behind the phenomena.
A significant question would be whether the transition temperature may become
even higher than 10 K by another type of chemical doping to these systems.
If so, such a study would not only find the highest ever magnetic transition
temperature in the Yb-based compounds at ambient pressure,
but it will also allows further in-depth study to elucidate
the electronic state change across the transition by spectroscopic method,
which has been otherwise impossible for Yb compounds to date. 

\begin{figure}[t!]
	\includegraphics[width=1\linewidth]{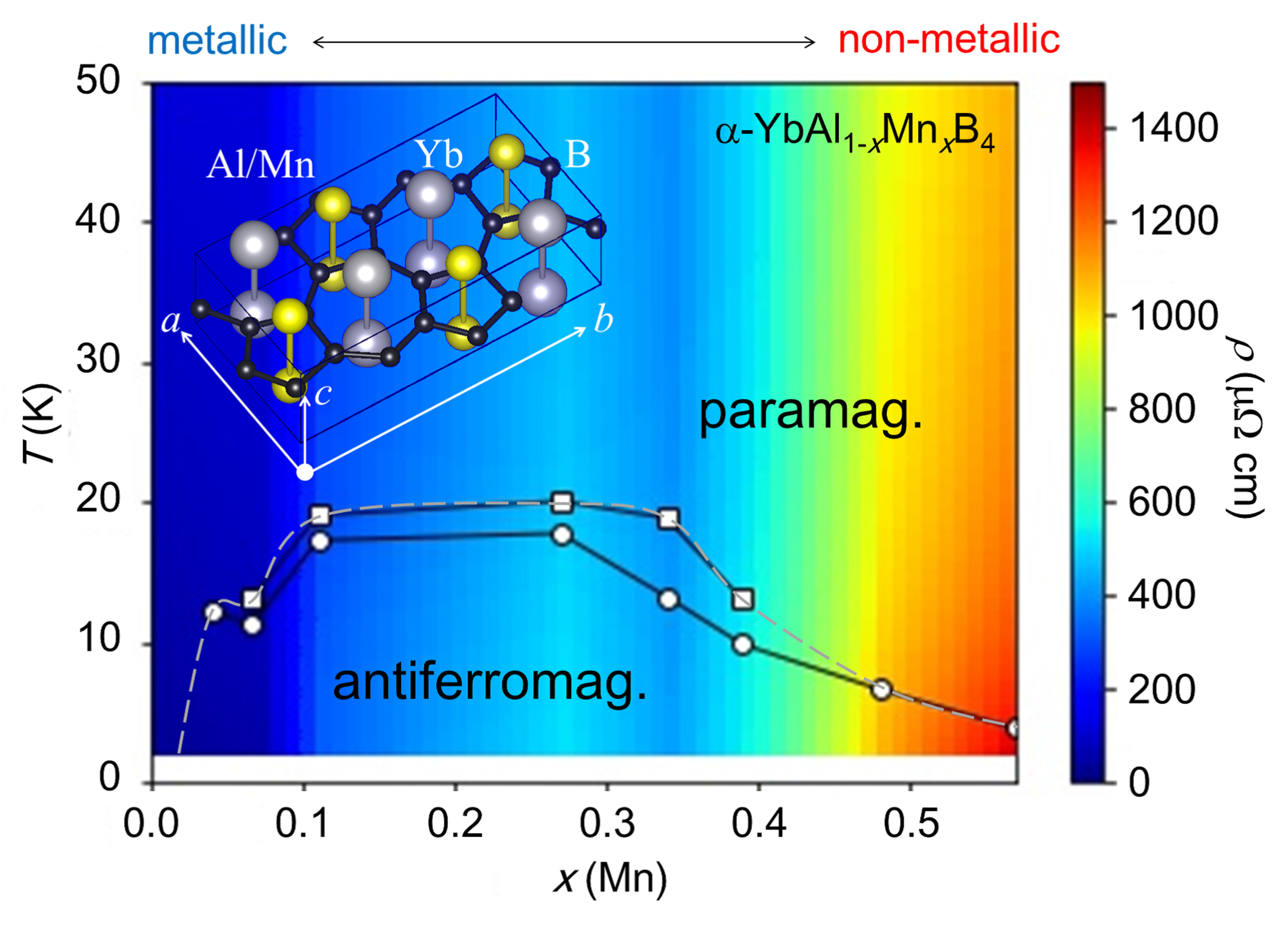}%
	\caption{\label{fig:phaseDiagram}
		(color online). 
		Phase diagram as a function of temperature $T$ (vertical axis)
                and Mn concentration $x$ (horizontal axis) for
                $\alpha$-YbAl$_{1-x}$Mn$_{x}$B$_{4}$.
		The contour plot of the $ab$-plane resistivity $\rho$ is also
                provided in the same region of $T$ and $x$.
		Each point indicates the N\'{e}el temperature $T_{\rm N}$
                determined by the anomaly found in the temperature dependence
                of the magnetization (squares) and specific heat (circles).
		The broken line schematically refers to an phase boundary due
                to the antiferromagnetic transition.
		Inset indicates the unit cell crystal structure of
                $\alpha$-YbAl$_{1-x}$Mn$_{x}$B$_{4}$.
	}
\end{figure}

Here, we report the discovery of a high N\'{e}el temperature reaching 20 K,
the highest temperature among the Yb based intermetallic compound at ambient
pressure.
As we show in Fig. \ref{fig:phaseDiagram}, the Mn substitution at the Al site
of 4\% is enough to induce an antiferromagnetic transition
in $\alpha$-YbAlB$_{4}$.
The transition temperature exceeds 20 K at the doping level of $\sim$ 30\% Mn.
The high transition temperature cannot be explained by the RKKY interaction
and thus points to an itinerant type magnetic order.
On the other hand, the combination of the transport and the photoemission
spectroscopy (PES) measurements reveals that the Mn doping induces
a hybridization gap at the Fermi energy ($E_{\rm F}$),
causing a highly resistive non-metallic behavior near $x({\rm Mn}) = 0.5$.
Our study indicates that the extremely high transition temperature appears
in a heavy fermion (HF) state proximate to a Kondo insulator phase.

\begin{figure*}[htb]
	\includegraphics[width=1\linewidth]{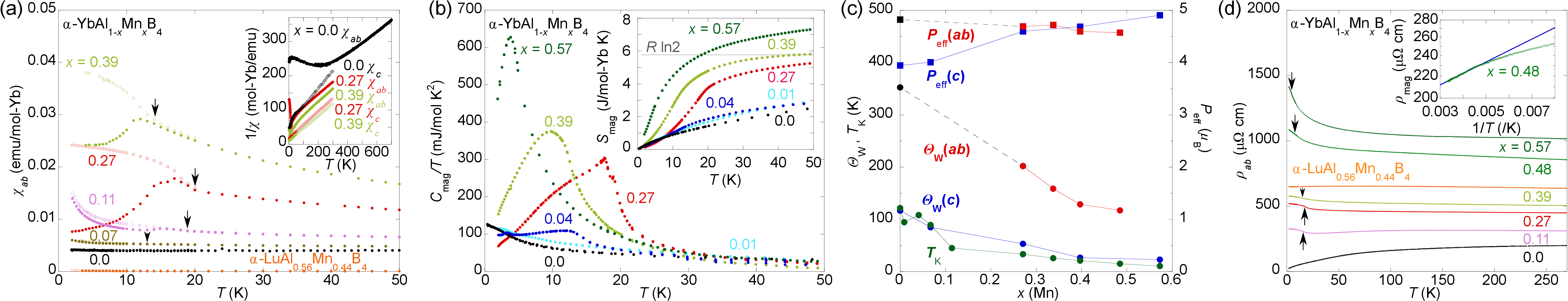}%
	\caption{\label{fig:characterization}
                (color online).
		(a) Temperature dependence of the magnetic susceptibility
                    measured in the magnetic field along the $ab$-plane
                    ($B$ = 1000 Oe).
		    Open circles and filled circles refer to the Field Cooled
                    (FC) and Zero Field Cooled (ZFC) components of the
                    susceptibility, respectively.
		    Arrows indicate the transition temperatures	defined
                    as the onset of the bifurcation between the FC and ZFC
                    results.
		    Inset shows temperature dependence of the inverse
                    susceptibility.
		(b) Temperature dependence of the magnetic part of the
                    specific heat divided by temperature, $C_{\rm mag}/T$. 
		    $C_{\rm mag}$ is estimated as
                    $C_{\rm mag} = C(\alpha$-YbAl$_{1-x}$Mn$_{x}$B$_{4})
                    -C(\alpha$-LuAlB$_{4})$.
		    Inset shows the temperature dependence of the entropy
                    calculated using the specific heat result.
		(c) Substitution $x$(Mn) dependence of
                    the Weiss temperature ($\Theta_{\rm W}$),
                    the Kondo temperature $T_{\rm K}$ (left axis)
                    and the effective moment ($P_{\rm eff}$) (right axis).
		(d) Temperature dependence of the $ab$-plane resistivity.
		    The inset indicates the semi logarithmic plot of the $1/T$
                    dependence of the magnetic part of the resistivity
                    $\rho_{\rm mag}$ for $x = 0.48$.
		    The magnetic part $\rho_{\rm mag}$ is defined as
                    $\rho_{\rm mag} =
                    \rho$($\alpha$-YbAl$_{0.52}$Mn$_{0.48}$B$_{4}$)
                    $-\rho$($\alpha$-LuAl$_{0.56}$Mn$_{0.44}$B$_{4}$).
		    \textcolor{black}{Black arrows indicate the transition
                    temperature estimated by the anomaly in the temperature
                    derivative of the resistivity and/or the peak of
                    the specific heat.
                    The solid line indicates the fit to the activation law
                    with the gap $\Delta = 50$ K (See text).
                    }
	}
\end{figure*}

Figures \ref{fig:characterization}(a) and \ref{fig:characterization}(b) display
the temperature dependence of the $ab$-plane magnetic susceptibility
and specific heat, respectively.
With increasing Mn concentration $x$, the temperature dependence of the
susceptibility exhibits a clear kink and a bifurcation between the zero field
cooled (ZFC) and field cooled (FC) components of the magnetic susceptibility.
While no magnetic anomaly is found at $x$ = 0.01,
the bifurcation is clearly seen at the concentration $x$ higher than 0.11,
and a small anomaly appears already at $x = 0.07$.
Correspondingly, the specific heat measurement finds a clear peak at almost
the same temperature as the bifurcation temperature of the magnetic
susceptibility curve.
The anomaly observed in both the susceptibility and specific heat demonstrates
that the antiferromagnetic transition is induced by the Mn substitution.
Strikingly, the transition temperature ($T_{\rm N}$) increases with increasing
chemical substitution and reaches a temperature as high as 20 K at $x = 0.27$.
On the other hand, the susceptibility measured for the nonmagnetic
analog $\alpha$-LuAl$_{0.56}$Mn$_{0.44}$B$_{4}$ has a much smaller value
around $\chi \sim 2 \times 10^{-4}$ emu/mol and its temperature dependence
does not exhibit any anomaly.
All these results including a $\mu$SR measurement
(given in the Supplemental Material \cite{supp})
indicate that Mn in $\alpha$-LuAl$_{0.56}$Mn$_{0.44}$B$_{4}$ should be
nonmagnetic.

To our knowledge, the antiferromagnetic transition temperature reaching 20 K
is the highest among the Yb-based magnets and HF compounds at ambient pressure.
Interestingly, $T_{\rm N}$ peaks at $x = 0.27$ and starts decreasing with
further substitution as found in both magnetic susceptibility and specific heat
results [Figs. \ref{fig:characterization}(a)
and \ref{fig:characterization}(b)].
It should be also noted that the Mn substitution changes the magnetic
anisotropy.
Whereas the pure $\alpha$-YbAlB$_{4}$ has the Ising type moment along
the $c$-axis and the $ab$-plane susceptibility is almost temperature
independent below 300 K,
the Mn substitution induces a more isotropic behavior with a strong temperature
dependence in the $ab$-plane,
as visible in the inverse susceptibility in the inset of
Fig. \ref{fig:characterization}(a).

To further understand the Mn substitution effects,
we fitted both the $ab$-plane and the $c$-axis components of
the susceptibility $\chi (T)$
using the Curie-Weiss (CW) law,
$\frac{1}{\chi} = \frac{3 k_{\rm B} \left( T - \Theta_{\rm W} \right)}
                       {N_{\rm A} P_{\rm eff}^{2} \mu_{\rm B}^{2} }$
where $k_{\rm B}$, $\Theta_{\rm W}$, $N_{\rm A}$, $P_{\rm eff}$
and $\mu_{\rm B}$ refer the Boltzmann constant, the Weiss temperature,
the Avogadro's number, the effective moment and the Bohr magneton,
respectively.
The fitting was performed in the temperature range from 150~K to 250~K
except for the $ab$-plane component of the pure system
[Fig. \ref{fig:characterization}(a) inset].
On the other hand, the magnetic part of the entropy $S_{\rm mag}$ was
estimated by
$S_{\rm mag} = \int \frac{C_{\rm mag}}{T} dT$, 
assuming a linear increase of $C_{\rm mag}$ between 0 and 2 K.
Here, $C_{\rm mag}$ indicates the magnetic part of the specific heat estimated
as noted in the caption of Fig. \ref{fig:characterization}(b).
Interestingly, the entropy in the paramagnetic state, e.g. at 40 K,
increases with increasing Mn substitution and reaches a value
of about $R\ln2$ for $x \ge 0.27$ [Fig. \ref{fig:characterization}(b) inset].
This indicates that the ground state remains a doublet,
but the associated Kondo temperature $T_{\rm K}$ decreases significantly
with increasing Mn substitution as we discuss in the following.

Figure \ref{fig:characterization}(c) shows the $x$(Mn) dependence of the Kondo
temperature $T_{\rm K}$, of the Weiss temperature $\Theta_{\rm W}$
and of the effective moment $P_{\rm eff}$.
$T_{\rm K}$ was estimated from the temperature dependence of the entropy
[Fig. 2(b) inset] using the relation 
$S_{\rm mag} \left( T = T_{\rm K} / 2 \right) = \left( R/2 \right) \ln 2$.
Both the Kondo temperature $T_{\rm K}$ and the Weiss
temperature $\Theta_{\rm W}(c)$, extracted from the $c$-axis component of
the susceptibility, nearly collapse with increasing $x$(Mn).
In its $ab$-plane component, the susceptibility of the pure system ($x=0$)
exhibits a peak at about 200$\sim$250 K due to the significant valence
fluctuations in this system \cite{matsumoto2011anisotropic,
PhysRevLett.104.247201,doi:10.7566/JPSJ.84.114715}.
Therefore, we extended our susceptibility measurements up to 700 K to
find a CW behavior above 300 K [Fig. 2(a) inset].
While both $T_{\rm K}$ and $\Theta_{\rm W} (c)$ are found to be of about
$\sim$ 130 K,
$\Theta_{\rm W} (ab)$ reaches about 350 K in the pure system from the fitting
of the $ab$-plane component of the susceptibility in between 600 K and 700 K. 
With increasing $x$(Mn), we find a systematic decrease in all
$\Theta_{\rm W} (ab)$, $\Theta_{\rm W} (c)$ and $T_{\rm K}$.
This indicates the valence fluctuation regime in $\alpha$-YbAlB$_{4}$ is
suppressed by the Mn substitution.   
In fact, the local moment $P_{\rm eff}$ gradually increases to reach
4.5 $\sim$ 4.7 $\mu_{B}$ at $x > 0.27$, a value equivalent to the one
of the isotropic Yb$^{3+}$ ion: 4.54 $\mu_{B}$.
Thus the Mn substitution leads the Yb ions to a trivalent configuration.

Generally, in the Kondo regime, the magnetic couplings are dominated by
the RKKY interaction.
Theoretically, the magnetic transition temperature due to the RKKY interaction
is known to be proportional to the de Gennes factor \cite{DeGennes1958}
(see the Supplemental Material \cite{supp}).
However, $T_{\rm N}$ of $\alpha$-YbAl$_{0.73}$Mn$_{0.27}$B$_{4}$ is strikingly
one order of magnitude higher than those with other $R$AlB$_{4}$ systems having
the YCrB$_{4}$ structure \cite{mori2007crystal,mori2011f}.
Namely, the magnetism in $\alpha$-YbAl$_{1-x}$Mn$_{x}$B$_{4}$ is not
a localized RKKY magnetism, but most likely an itinerant type.

\textcolor{black}{
	An important effect to understand the origin of the itinerant magnetism
        is the increase in the hybridization strength with the Mn substitution.
        The substitution decreases the volume of the unit cell \cite{supp},
        and thus applies chemical pressure to the system.
	The valence change discussed above from the larger Yb$^{2+}$ to
        smaller Yb$^{3+}$ ions also causes a similar trend.   
	Therefore, the bond lengths between Yb and B become shorter, and thus
	the hybridization strength should increase
        with the Mn substitution \cite{Goltsev2005,Flouquet2012}.
	Further evidence for the itinerant magnetism will be discussed below
        using the results obtained in the resistivity and PES measurements.
}

Recent structure analysis using powder X-ray diffraction under high
pressure on $\beta$-YbAlB$_{4}$ indicates that the B rings sandwiching Yb ion
(Fig. \ref{fig:phaseDiagram} inset)
lowers its symmetry from nearly 7 fold by application
of pressure \cite{sakaguchi2016pressure}.
One theoretical approach pointed out that this 7 fold symmetry stabilizes
the Yb ground state to $|J_{z} = \pm 5/2>$.
This orbital character is shown to imply an anisotropic hybridization
as well as the Ising behavior of the magnetic moment along the $c$-axis
at ambient pressure \cite{nevidomskyy2009layered,ramires2012beta}. 
These results suggest that the deformation of the local symmetry under pressure
suppresses the Ising-like behavior and induces the 30 K transition
under 8 GPa in $\beta$-YbAlB$_{4}$.

This may be also the case in $\alpha$-YbAlB$_{4}$ and the magnetism induced
by the Mn substitution.
Nearly the same temperature dependence of the susceptibility has been seen
for both phases of YbAlB$_{4}$ at ambient pressure.
Thus, the crystal electric field (CEF) scheme must be nearly the same for both
phases \cite{matsumoto2011anisotropic}.
Indeed, a recent measurement of X-ray magnetic circular dichroism
indicates the ground state of $\alpha$-YbAlB$_{4}$ to be
the $|J_{z} = \pm 5/2>$ state \cite{doi:10.7566/JPSJ.84.114715}.
On the other hand, the systematic change in anisotropy of the susceptibility
as visible in the inset of Fig. \ref{fig:characterization}(a) and
in Fig. \ref{fig:characterization}(c) demonstrates that the CEF ground state
changes with the Mn substitution, while keeping a doublet configuration
as the magnetic entropy at 40 K remains around $R\ln2$.
Thus, most likely the ground doublet departs
from the $\left| J_{z} = \pm 5/2 \right \rangle$ state due to the local
symmetry deformation around the Yb-site, itself induced by the Mn substitution.
This change of ground-state plays an important role in the high temperature
\textcolor{black}{magnetism, as discussed below. }

Another effect of the chemical substitution is the carrier doping.
Such carrier doping may well change the transport properties, and therefore,
we performed the longitudinal resistivity measurements
[Fig. \ref{fig:characterization}(d)].
The magnitude as well as the temperature dependence of the resistivity
dramatically change with the Mn substitution.
The good metallic behavior observed in the pure system is already lost
at $x = 0.11$ and the resistivity at both room temperature and 2 K increases
nearly linearly with $x$.
The resistivity at 2 K for $x = 0.57$ is $\sim$ 60 times larger than the one
for $x = 0$.
Moreover, the temperature dependence systematically changes into a more
incoherent semimetallic or non-metallic behavior with Mn doping.
These indicate that the carrier concentration must decrease with
increasing Mn concentration.
We also point out that the high temperature resistivity for $x \sim 0.5$ can
be roughly fitted with the activation law, namely,
$\rho = \rho_0 \exp (\Delta/T)$ with an activation energy of
about $\Delta \sim 50$ K.
The inset of Fig. \ref{fig:characterization}(d) exemplifies such a fit with
the magnetic part of the resistivity for $x = 0.48$ in which the activation law
is observed at high temperature from 200 K to above 300 K.
This suggests that the highly resistive transport induced by Mn substitution
results from the system approaching an insulating state.
The deviation from the activation law at low temperature indicates
the presence of impurity bands which cause the saturation of the resistivity
on cooling.
\textcolor{black}{
In addition, a clear anomaly at $T_{\rm N}$ can be found for
the high $T_{\rm N}$ samples with $x$ = 0.11 and 0.27.
Ce(Ru$_{0.85}$Rh$_{0.15}$)$_2$Si$_2$ exhibits a quite similar behavior
at a spin-density-wave (SDW) transition \cite{Nakano2003}.
This suggests the SDW transition as the origin of the anomaly seen
between 0.04 $\leqq$ $x$ $\leqq$ 0.27.
}

\begin{figure}[htb]
	\includegraphics[width=1\linewidth]{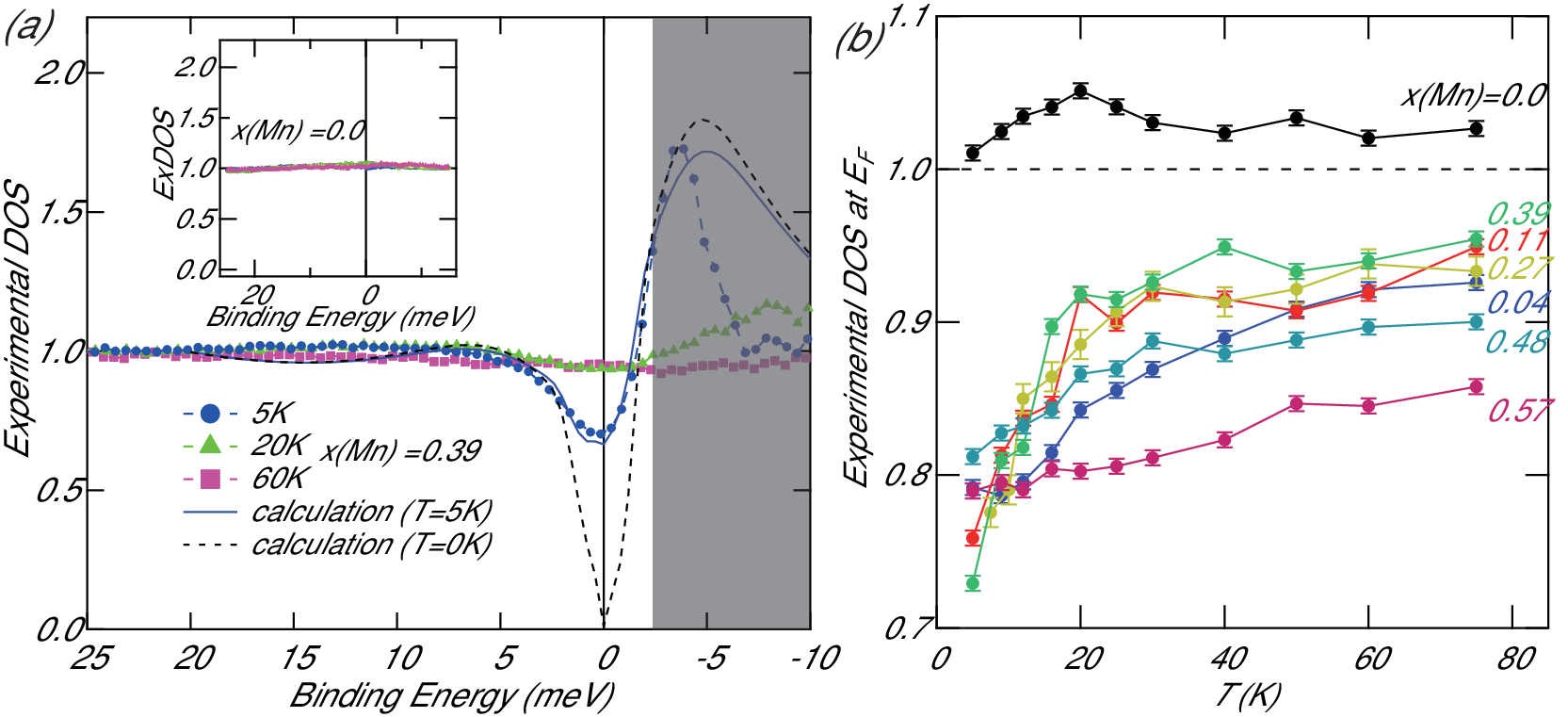}%
	\caption{\label{fig:PES}
                 (color online).
                 (a) Experimental density of state (ExDOS) near $E_{\rm F}$
                     for $\alpha$-YbAl$_{1-x}$Mn$_{x}$B$_{4}$ ($x$ = 0.39)
                     derived from the PES measurements.
                     The dotted lines indicate the calculation based on the
                     \textcolor{black}{periodic Anderson model} \cite{supp}.
		     \textcolor{black}{The uncertain energy region on ExDOS
                     over $\sim$$5k_{\rm B}T$ above
                     $E_{\rm F}$ \cite{Ehm2007,supp} is shaded
                     (over $\sim$2.2 meV at 5 K). }  
		     Inset shows ExDOS for $x$ = 0.0. ExDOS for other $x$ is
                     displayed in the Supplemental Material \cite{supp}.
		 \textcolor{black}{(b)} Temperature dependence of ExDOS
                     at $E_{\rm F}$ for various compositions. 
	}
\end{figure}

To investigate the origin of the insulating behavior, 
\textcolor{black}{
the density of states (DOS) near $E_{\rm F}$ was examined using
the high-resolution and bulk-sensitive
laser PES \cite{kiss2005photoemission,shimojima2015low}.}
In order to evaluate the DOS below and slightly above $E_{\rm F}$,
the PES spectra were divided by spectra measured on a gold reference
at each measured temperature.
We should stress that this method is accurate only up to 5$k_{\rm B}$T
above $E_{\rm F}$ \cite{Ehm2007}, as indicated by the shaded energy range
of Fig. \ref{fig:PES}(a).
The derived results, referred as experimental DOS (ExDOS), scarcely depends
on the temperature for $x = 0$ as shown by the inset of Fig. \ref{fig:PES}(a).
On the other hand, for $x = 0.39$, ExDOS near $E_{\rm F}$ is mostly flat
at $T$ = 75 K and gets suppressed as the temperature decreases.
For quantitative discussion, ExDOS at $E_{\rm F}$ is plotted for each measured
doping in Fig. \ref{fig:PES}(b) as a function of temperature.
For $x = 0$, ExDOS at $E_{\rm F}$ exceeds 1, indicating a metallic state,
in agreement with previous angle resolved PES
measurements \cite{PhysRevB.97.045112}.
In contrast, ExDOS at $E_{\rm F}$ for the doped samples ($x > 0$) falls below 1
and decreases when the temperature is lowered.
The temperature-dependent depression of ExDOS is observed for all
of the doped samples and starts \textcolor{black}{far} above the magnetic
transition temperature $T_{\rm N}$,
although the spectral change appears larger around
\textcolor{black}{$T\sim$ 15 K}. 
These results are consistent with the resistivity measurements and indicate
the formation of a pseudo-gap, or a $c$-$f$ hybridization gap, for $x > 0$.
One may notice that, for $x$ = 0.39, ExDOS across $E_{\rm F}$ is highly
asymmetric and finite ($>$ 0) at $E_F$ even at $T$ = 5 K.
A theoretical calculation based on the
\textcolor{black}{periodic Anderson model} with $c$-$f$ hybridized
bands \cite{Hewson1993,Generalov2017}
can reproduce such DOS near $E_{\rm F}$ at $T$ = 5 K as also plotted
in Fig. \ref{fig:PES}(a) \cite{supp}.
In addition, DOS at $E_{\rm F}$ reaches 0 in the calculation for $T$ = 0 K
using the same parameters set for $T$ = 5 K,
without any experimental broadening factor.
This theoretical modeling suggests the gap formation observed with Mn doping
to emerge from a Kondo-like hybridization.
Since the gap opens over $E_{\rm F}$, it naturally relates to the systematic
increase of resistivity and pictures a crossover from an intermediate valence
metal to a Kondo insulator as Al is substituted with Mn. 
\textcolor{black}{
Such Kondo-like hybridization further implies that the Yb 4$f$-electron bands
around $E_{\rm F}$ directly contribute to the transport
and thus to the itinerant magnetism of $\alpha$-YbAl$_{1-x}$Mn$_{x}$B$_{4}$.} 
The spectral shape and gap formation observed
in ExDOS are further discussed in the Supplemental Material \cite{supp}.

While Kondo insulators are commonly found in cubic systems,
some orthorhombic systems, of symmetry similar to the present one,
are known to be Kondo insulators \cite{takabatake1990formation,
malik1991evidence,kumigashira1999spectral,muro2010structural,
nishioka2009novel,muro2010structural,takesaka2010semiconducting}.
In particular, Ce$Tr_{2}$Al$_{10}$ ($Tr$ = Os, Ru) also shows a high magnetic
transition temperature which motivated intensive studies to reveal
its origin \cite{kimura2011electronic,kawabata2014suppression}.
The similar highly resistive state, proximate to a Kondo insulator, found
in $\alpha$-YbAl$_{1-x}$Mn$_{x}$B$_{4}$ suggests that the same type of
mechanism is at the origin of the high temperature magnetism.

\textcolor{black}{
Overall results indicate that the high temperature magnetism in this system
is the itinerant SDW-type induced by chemical substitution,
similar to the Rh substitution case in CeRu$_2$Si$_2$ \cite{Nakano2003}.
The hybridization strength increases with the Mn substitution,
while the suppression of the valence fluctuations in the Kondo regime
-- with an almost trivalent Yb configuration --
leads to the decreases of the Kondo and Weiss temperatures.
Theoretically, the strong hybridization may lead to a SDW-type order
near the onset of a Kondo insulating phase \cite{Yoshida2011}.
The sharp coherent \textcolor{black}{feature} in PES seen only
for the high $T_{\rm N}$ samples between $x = 0.11$ and 0.39 \cite{supp}
may result from the nesting across $E_{\rm F}$ leading to a SDW-order.
}

\textcolor{black}{
The itinerant 4$f$-electrons in pure $\alpha$-YbAlB$_{4}$ form quasi-two
dimensional bands. They do not contribute much to transport along
the $c$-axis but do mainly in the $ab$-plane \cite{matsumoto2011anisotropic}.
This suggests that the SDW-type antiferromagnetism is formed
with the nesting vector in the $ab$-plane.
The suppression of the magnetic anisotropy found in the susceptibility
measurements by Mn doping indicates that the CEF, which initially
implies the Ising-like ground state along the $c$-axis of the pure system,
is modified by the distortion of the 7-fold local symmetry at the Yb site.
As a result, the 4$f$ magnetic moment may develop a component
in the $ab$-plane.
As well known for the spin-flop transition in Cr \cite{PhysRev.155.528},
the SDW energy scale can be affected by the spin configuration 
and by its nesting vector.
Therefore, the high $T_{\rm N}$ induced by Mn substitution might also arise
from the change in the CEF scheme.
Future studies of the CEF scheme, both experimental
(e.g. neutron scattering, NMR measurement) and theoretical,
are necessary to clarify such possible mechanism.
}

To conclude, through the comprehensive measurements
of $\alpha$-YbAl$_{1-x}$Mn$_{x}$B$_{4}$, we found that
the Mn substitution induces a high temperature antiferromagnetism,
whose transition temperature reaches 20 K at $x = 0.27$.
This is so far the highest magnetic transition temperature among
the Yb based HF systems at ambient pressure.
Our transport measurements show that the system exhibits a high resistivity
with the Mn substitution.
Thanks to PES, we further highlight the formation of a gap at the Fermi level,
together with a coherent feature, consistent with the resistivity measurements.
These observations evidence that the Mn substitution results in an increase
of the Kondo-like hybridization, driving the system out of the mixed valence
state of the pure $\alpha$-YbAlB$_{4}$ to a more localized Kondo HF state.
Additionally, the onset of the gap formation in the PES measurements appears
at temperature well above the magnetic transition temperature;
it lifts any ambiguity about the origin of the gap and underlines the crossover
from a heavy Fermi liquid to a Kondo insulator with increasing Mn doping. 

\begin{acknowledgments}
\section*{Acknowledgements}
We thank K. Sone, C. Broholm, S. Wu, M. Suzuki, R. Arita for useful
discussions.
This work is partially supported by CREST (JPMJCR18T3), PRESTO (JPMJPR15N5),
Japan Science and Technology Agency, by Grants-in-Aids for Scientific Research on Innovative Areas (JP15H05882 and JP15H05883) from the Ministry of Education, Culture, Sports, Science, and Technology of Japan, 
and by Grants-in-Aid for Scientific Research (JP25707030, JP25887015, JP15J08663, JP16H02209, JP16H06345, JP19H00650) from the Japanese Society for the Promotion of Science (JSPS).
The use of the facilities of the Materials Design and Characterization Laboratory at the Institute for Solid State Physics, The University of Tokyo, is acknowledged.
S. Suzuki was supported by Japan Society for the Promotion of Science
through Program for Leading Graduate Schools (MERIT).
This research is funded in part by a QuantEmX grant from ICAM,
the Gordon and Betty Moore Foundation through Grant GBMF5305 and
by Canadian Institute for Advanced Research.
S. Suzuki greatly appreciates the hospitality of Department of Physics
and Astronomy of Johns Hopkins University,
where a part of this work was conducted.
Institute for Quantum Matter, an Energy Frontier Research Center was funded by DOE, Office of Science, Basic Energy Sciences under Award \# DE-SC0019331.
\end{acknowledgments}


%
%


\end{document}